\documentclass[%
 aps,
 prl,%
 amsmath,amssymb,
reprint,%
superscriptaddress,showpacs]{revtex4-1}

\usepackage{color, graphicx}
\usepackage{dcolumn}
\usepackage{bm}
\usepackage{amssymb}
\usepackage{latexsym}
\usepackage{amsfonts}
\usepackage{amsmath}
\usepackage{multirow}
\usepackage{ifthen}
 \usepackage{ulem}

\begin{document}

\title{Unidirectional zero sonic reflection in passive $\mathcal{PT}$ symmetric Willis media}

\author{Aur\'elien Merkel}
\affiliation{Department of Physics, Universidad Carlos III de Madrid, Avenidad de la Universidad 30, 28911 Legan\'{e}s (Madrid), Spain}

 \author{Vicent Romero-Garc\'ia}
 \affiliation{Laboratoire de l'Universit\'e du Mans, LAUM UMR-6613 CNRS, Le Mans Universit\'e, Avenue Olivier Messiaen, 72085 Le Mans cedex 9, France}
 
 \author{Jean-Philippe Groby}
 \affiliation{Laboratoire de l'Universit\'e du Mans, LAUM UMR-6613 CNRS, Le Mans Universit\'e, Avenue Olivier Messiaen, 72085 Le Mans cedex 9, France}

\author{Jensen Li}
\affiliation{Department of Physics, The Hong Kong University of Science and Technology, Hong Kong, China}

\author{Johan Christensen}
\affiliation{Department of Physics, Universidad Carlos III de Madrid, Avenidad de la Universidad 30, 28911 Legan\'{e}s (Madrid), Spain}


\begin{abstract}
In an effective medium description of acoustic metamaterials, the Willis coupling plays the same role as the bianisotropy in electromagnetism. Willis media can be described by a constitutive matrix composed of the classical effective bulk modulus and density  and additional cross-coupling terms defining the acoustic bianisotropy. Based on an unifying theoretical model, we unite the properties of acoustic Willis coupling with $\mathcal{PT}$ symmetric systems under the same umbrella and show in either case that an exceptional point hosts a remarkably pronounced scattering asymmetry that is accompanied by one-way zero reflection for sound waves. The analytical treatment is backed up by experimental input in asymmetrically side-loaded wavesguides showing how gauge transformations and loss biasing can embrace both Willis materials and non-Hermitian physics to tailor unidirectional reflectionless acoustics, which is appealing for purposeful sound insulation and steering. 
\end{abstract}

\maketitle
Metamaterials and metasurfaces have emerged as artificial structures that enable one to tailor the wave propagation at will \cite{WegenerScience}. Whether the field of interest is optics, acoustics, or elasticity, recent efforts have enabled exiting wave phenomena ranging from cloaks of invisibility and unhearability, zero-index behaviour and negative refraction, just to name a few of many counterintuitive effects \cite{milton2006,norris2008,torrent2008,garciachocano2014,haberman2016}. Wave-based diodes and rectifiers enabling one-way sound or light propagation have recently also been put on the map of contemporary metamaterial research \cite{liang2010-2,boechler2011,fleury2014,devaux2015,trainiti2016,merkel2018prapp}.\\

Waves irradiating onto a passive reciprocal medium usually display symmetric transparency in that the transmittance does not depend on the side at which waves are launched. In contrast, if the elementary building units of the medium involved lack intrinsic inversion symmetry, then the reflectance depends on from which side of the slab waves are irradiated. Commonly, for electromagnetic (EM) waves, such asymmetric response are known to occur in bianisotropic media, which are important in applications comprising unidirectional radiation, single-sided light detection and emission \cite{cheng1968,kong1972,Kriegler,marques2002,liangpeng2018}. The intrinsic asymmetry of the involved meta-atoms leads to subwavelength cross-coupling where magnetic states can be excited also by electric fields, and likewise, electric states that can be excited through magnetoelectric coupling. An analogue picture for the case of sound is provided through the cross-coupling between strain and velocity in so-called \textit{Willis media} \cite{WillisSeminal,milton2007}. Local and non-local coupling in the form of microstructural asymmetry and finite phase changes across inhomogeneities, respectively, define the very nature of Willis coupling in bianisotropic artificial acoustic or mechanical media \cite{PhysRevB.92.174110,koo2016,muhlestein2016,muhlestein2017,sieck2017,liquan2018}. To this extent, Willis media have not only emerged to obtain physically meaningful effective parameters, but also for the design of passive acoustic one-way reflectionless systems with the potential to engineer asymmetric steering of sound and to create mechanical structures that suppress echoes from one side, but act reflective from the other. \\

Asymmetries related to wave propagation have been made also possible in systems disobeying parity $\mathcal{P}$ and time-reversal $\mathcal{T}$ symmetry, but that are nevertheless symmetric under simultaneous operation. $\mathcal{PT}$ symmetry or non-Hermiticity in optics or acoustics, embraces the introduction of loss and gain constituents in new synthetic material such as metamaterials. In optics, gain is an essential ingredient to overcome metamaterial losses but has found increasing importance in one-way waveguides, single-sided scattering, and enhanced sensing at the exceptional point (EP), which marks the transition between an exact and a broken $\mathcal{PT}$-symmetry phase \cite{elganainy2018}. Specifically, among the phenomena occurring at the EP, unidirectional reflectionless wave propagation, which is also called anisotropic transmission resonance, originates from a non-Hermitian degeneracy of the scattering matrix \cite{zinlin2011,lige2012PRA,fleury2015,chengzhishi2016,PhysRevLett.116.207601}. Interestingly, unidirectional zero reflection has also been found in entirely passive non-Hermitian systems without involving gain, thus posing significant ease for experimental realizations \cite{Aguo2009,ruter2010,minkang2013,liangfeng2013,yongsun2014,lawrence2014,jinhuiwu2014,yixinyan2014,tuoliu2018}.  \\ 

In this letter we present a theoretical unification framework that merges the properties of the acoustic constitutive relations containing local and nonlocal Willis coupling with a conventional $\mathcal{PT}$ symmetric system representation at the EP in asymmetrically side-loaded waveguides. In details, based on experimental data, we map the one-way reflectionless sound propagation in Willis media onto an ideal, that is, a gain and loss balanced $\mathcal{PT}$ symmetric system through a gauge transformation comprising a shift in the $\mathcal{P}$ operation associated with an average loss bias. This unconventional strategy to tailor the asymmetric flow of sound further extends the family of non-Hermitian wave physics to synthesized $\mathcal{PT}$ matter.\\

We begin by considering an effective medium containing Willis coupling, hence, the conservation of momentum in an acoustic waveguide is 
\begin{equation}
\nabla P=-\frac{\rho_{\text{eff}}}{\rho_0c_0S} \frac{\partial \textbf{U}}{\partial t} - \boldsymbol{\chi}_1\frac{\partial P}{ \partial t}, 
\end{equation}
and the conservation of mass reads 
\begin{equation}
\nabla\cdot \textbf{U}=-\frac{\rho_0c_0 S} {K_{\text{eff}}}\frac{\partial P}{ \partial t} - \boldsymbol{\chi}_2 \cdot \frac{\partial \textbf{U}}{\partial t},  
\end{equation}
where $P\equiv P/(\rho_0c_0)$ is the acoustic pressure, $\rho_0$ is the mass density and $c_0$ is the speed of sound in the background medium, $\textbf{U}\equiv S\textbf{U}$ is the acoustic volume flow and $S$ is the cross-section of the cylindrical waveguide, $ \boldsymbol{\chi}_{1,2}$ are the Willis coupling parameters, and $\rho_{\text{eff}}$, $K_{\text{eff}}$ represent the effective mass density and bulk modulus, respectively. 
Because of reciprocity, the relation $-\boldsymbol{\chi}_1=\boldsymbol{\chi}_2\equiv \boldsymbol{\chi}$ is always satisfied. The constitutive matrix $\textbf{M}$ in one dimension along the $z$ axis with the time convention  $e^{-i \omega t}$,  has the following form 
\begin{eqnarray}
\frac{\partial}{\partial z}\Pi\left(\begin{array}{c} P\\U \end{array}\right)
=i\omega\left[\begin{array}{cc} S\rho_0c_0/K_{\text{eff}} & \chi \\  -\chi & \rho_{\text{eff}}/(S\rho_0c_0) \end{array}\right]
\left(\begin{array}{c} P \\U\end{array}\right), 
\label{eqconstitutiveAc}
\end{eqnarray}
with the operator
\begin{eqnarray}
\Pi=\left(\begin{array}{cc} 0 & 1 \\ 1 & 0\end{array}\right).
\end{eqnarray}

\begin{figure}[ht!]
\centering
\includegraphics[width=8.6cm]{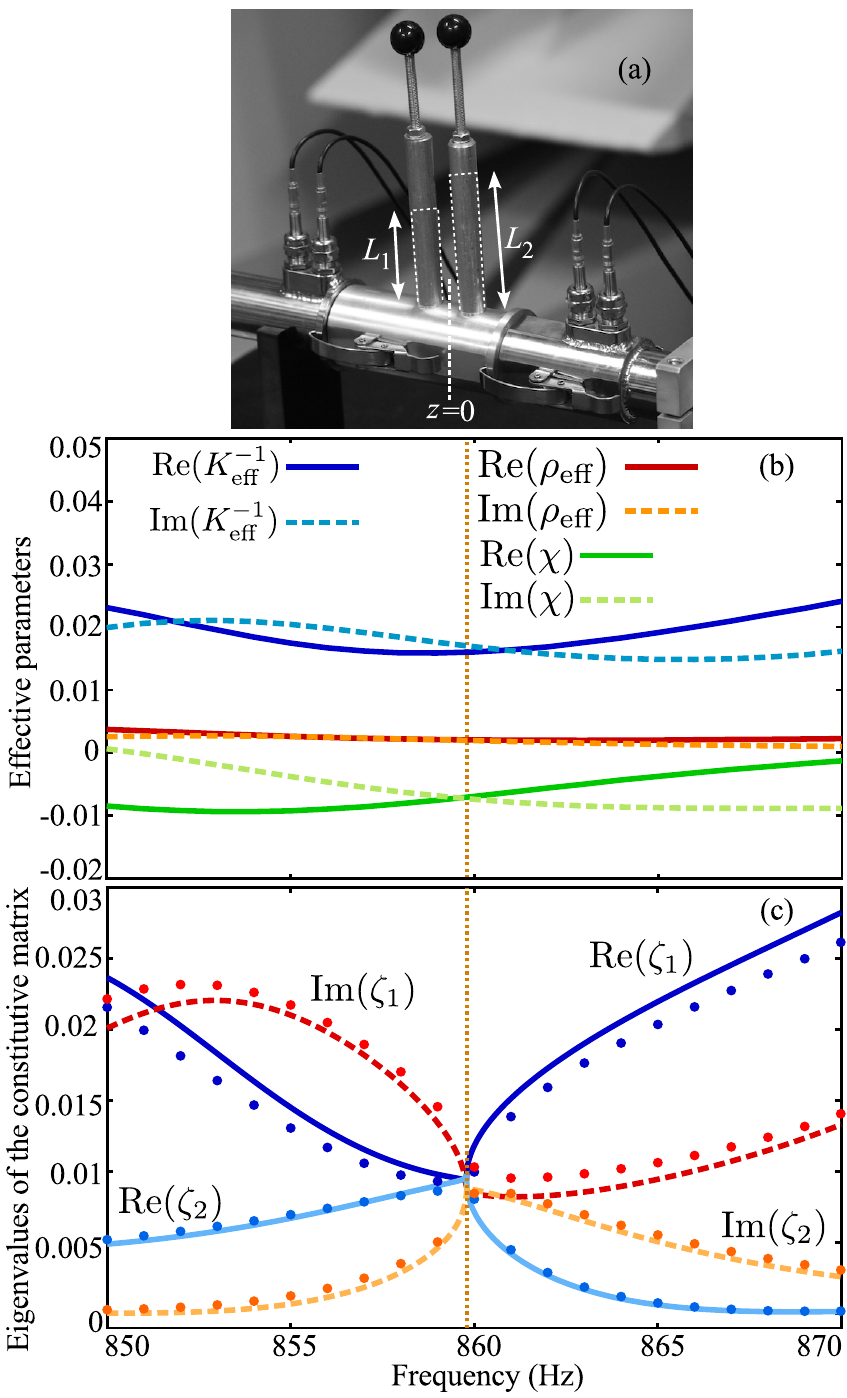}
\caption{(color online) (a) Picture of the experimental setup with the acoustic waveguide side-loaded by two quarter-wavelength resonators of different resonance frequencies. (b) Experimentally derived effective parameters $K_{\text{eff}}$, $\rho_{\text{eff}}$ and $\chi$. (c) Theoretical (continuous curves) and experimental (dots) eigenvalues $\zeta_{1,2}$ of the constitutive matrix $\textbf{M}_{\textbf{eff}}$. The EP where the eigenvalues coalesce is marked with an orange vertical line. \label{figeigH}}
\end{figure}

We can find the elements of the constitutive matrix from the scattering coefficients, which can be derived from the transfer matrix method \cite{merkel2015APL,jimenez2017}. The acoustic pressure and velocity located before and after the cell of length $L$ can be related via the constitutive matrix $\textbf{M}$ as follows
\begin{eqnarray}
\left[\begin{array}{c} P_{z=L} \\ U_{z=L}\end{array}\right]=e^{i\omega \Pi \textbf{M}L}\left[\begin{array}{c} P_{z=0} \\ U_{z=0}\end{array}\right],
\end{eqnarray}
leading to
\begin{eqnarray}
e^{i\omega \Pi \textbf{M}L} &=&
\left[\begin{array}{cc}  1/(\rho_0c_0) & 1/(\rho_0c_0) \\ 1/(SZ_w) & -1/(SZ_w) \end{array}\right]
\left(\begin{array}{cc} t-r_Lr_R/t & r_R/t \\ -r_L/t & 1/t\end{array}\right) \nonumber \\
& &\times \left[\begin{array}{cc} 1/(\rho_0c_0) & 1/(\rho_0c_0) \\1/(SZ_w) & -1/(SZ_w) \end{array}\right]^{-1}, 
\label{eqinversion}
\end{eqnarray}
where $t$ is the complex and reciprocal transmission coefficient, $r_{L,R}$ are the complex reflection coefficients from left $L$ and right $R$ incidence, respectively, and $Z_w$ is the acoustic impedance of the waveguide. In Eq.~(\ref{eqconstitutiveAc}) we can refrain from normalizing the acoustic parameters to obtain an effective constitutive matrix
\begin{eqnarray}
\textbf{M}_{\textbf{eff}}=\left(\begin{array}{cc} K_{\text{eff}}^{-1} & \chi \\  -\chi & \rho_{\text{eff}} \end{array}\right),
\label{eqMeff}
\end{eqnarray}
which has the eigenvalues $\zeta_{1,2}$. In order to realize such effective metamaterial system containing an asymmetric subwavelength cell, we fabricated an acoustic waveguide with two side-loaded resonators tuned at two distinct resonance frequencies as depicted in Fig.~\ref{figeigH}(a). We take the case of two quarter-wavelength resonators of radius $R_{1,2}=5.35$~mm having their resonance frequencies at $f_{1}=895$~Hz and $f_{2}=862$~Hz, corresponding to the lengths  $L_{1}=9.141$~cm and $L_{2}=9.51$~cm, respectively and separated by a distance $L=3$~cm in the waveguide of radius $R_w=1.5$~cm. The scattering coefficients are measured in an impedance tube with the four microphone method. The effective parameters, which are derived from the matrix logarithm of Eq. (\ref{eqinversion}), exhibit nonzero Willis coupling as can be seen in Fig.~\ref{figeigH}(b). Whereas peculiar features do not appear obvious in $\rho_{\text{eff}}$ and $K_{\text{eff}}$, the eigenvalues $\zeta_{1,2}$ coalesce at approximately 860~Hz as can be seen in Fig.~\ref{figeigH}(c) demonstrating the formation of an EP. At this point, we show that the constitutive matrix $\textbf{M}_{\text{eff}}$ can be mapped onto a conventional 2$\times$2 $\mathcal{PT}$ symmetric system matrix of the form \cite{bender2002,mostafazadeh2003,minkang2013,kunding2016}
\begin{eqnarray}
\Upsilon_{PT} = 
\left(\begin{array}{cc}
a+i\Delta \gamma & i\kappa \\
-i\kappa & a-i\Delta \gamma
\end{array}\right),
\label{eqPTconv}
\end{eqnarray}
where $a$, $\Delta\gamma$ and $\kappa$ are real numbers. Similarly, $\textbf{M}_{\text{eff}}$ can also be mapped onto an entirely real 2$\times$2 matrix that has the form
\begin{eqnarray}
\Upsilon_R  & = & \left(\begin{array}{cc}
a+\Delta\gamma & \kappa \\
-\kappa & a-\Delta\gamma
\end{array}\right).  
\label{eq_realantisym}
\end{eqnarray}
These two $\mathcal{PT}$-symmetric two-level matrices have an EP occurring when $\Delta \gamma = \pm \kappa$. 
\begin{figure}[ht!]
\centering
\includegraphics[width=8.6cm]{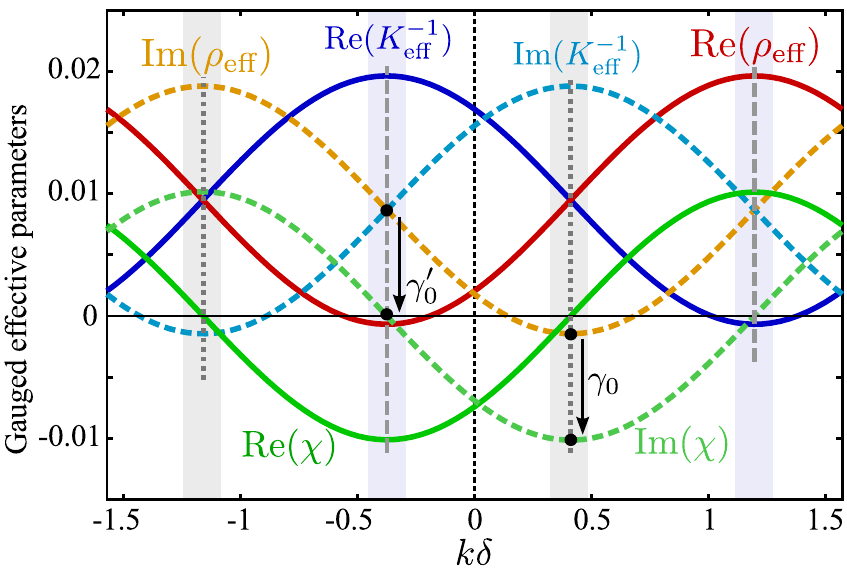}
\caption{(color online) Gauge transformation of Eq. (\ref{eq_unitygauge}) comprising the elements of the constitutive matrix $\textbf{M}_{\textbf{eff}}$ at the EP. The gauge transformation maps the constitutive matrix onto a $\mathcal{PT}$-symmetric matrix $\Upsilon_{PT}+i\gamma_0\textbf{I}$, where $\text{Re}(\chi)=0$, marked with vertical gray dotted lines in the gray areas or onto a real matrix $\Upsilon_{R}+i\gamma'_0\textbf{I}$, where $\text{Im}(\chi)=0$, marked with vertical gray dashed lines in the blue areas. \label{figGauge}}
\end{figure}
At the EP of our system, $\text{Re}(K^{-1}_{\text{eff}})\neq\text{Re}(\rho_{\text{eff}})$,  $\text{Im}(K^{-1}_{\text{eff}})\neq-\text{Im}(\rho_{\text{eff}})$ and $\chi$ is complex as can be seen in Fig. \ref{figeigH}(b), hence, at first sight, no form similarity seems to exist between $\textbf{M}_{\textbf{eff}}$, $\Upsilon_{PT}$ or $\Upsilon_R$. However, the constitutive matrix $\textbf{M}_{\textbf{eff}}$ can be gauged through an unitary transformation \cite{gear2017}
\begin{equation}
\textbf{M}_{G} = U_u \textbf{M}_{\textbf{eff}}U_u^{-1}, \label{eq_unitygauge}
\end{equation}
where $U_u$ is the propagator matrix
\begin{eqnarray}
U_u=
 \left(\begin{array}{cc}
 \cos(k\delta) & i \sin(k\delta) \\
i \sin(k\delta) & \cos(k \delta)
\end{array}\right), 
\end{eqnarray}
in which $k$ is the wavenumber and $\delta$ is real. The matrix $U_u$ is unitary, i.e., $U_u^{-1}=U_u^{\dagger}$, ensuring the eigenvalues of any transformed matrix $\textbf{M}_G$ remain invariant. This unitary gauge transformation maps ideal $\mathcal{PT}$ symmetric systems such as $\Upsilon_{PT}$ and $\Upsilon_{R}$ of symmetric $\mathcal{P}$ operation, that is, $z\rightarrow -z$, which is a mirror operation at the center of the cell onto shifted $\mathcal{PT}$ symmetric systems accommodating the generalized $\mathcal{P}$ operation $z\rightarrow 2\delta-z$, which is a mirror operation at the position $\delta$ from the center of the cell \cite{cannata2007}. We begin by gauging the constitutive matrix $\textbf{M}_{\textbf{eff}}$ through the dimensionless factor $k\delta$ in order to obtain fully imaginary Willis coupling, $\text{Re}(\chi)=0$, as highlighted with vertical dotted lines in the gray areas in Fig.~\ref{figGauge}, which is accompanied by two exact analytical constitutive relationships $\text{Re}(K_{\text{eff}}^{-1})=\text{Re}(\rho_{\text{eff}})$ and $\text{Im}(K_{\text{eff}}^{-1}-\rho_{eff})=\pm 2 \text{Im}(\chi)$. The resulting gauged constitutive matrix then obtains $\pm\text{Im}(\chi)$ in the off-diagonal terms whereas the real parts of the diagonal terms become $\text{Re}(K_{eff}^{-1})$. Since $\Upsilon_{PT}$ represents an ideal balance of loss and gain, $\textbf{M}_{\textbf{eff}}$ cannot exactly assume the same form. However, in order to accomplish this, we introduce an additional gauge transformation that is biasing the system with an average level of loss $\gamma_0$ \cite{Aguo2009,ruter2010,minkang2013,liangfeng2013,yongsun2014,lawrence2014,jinhuiwu2014,yixinyan2014,tuoliu2018}. In doing this, the ideal $\mathcal{PT}$-symmetric Hamiltonian is mapped onto an effective but passive Hamiltonian that yields a loss-biased constitutive matrix of the following form
\begin{eqnarray}
\textbf{M}_{\textbf{eff}}  & = & 
 U_u \left(\begin{array}{cc}
a+i\Delta \gamma+i \gamma_0 & i\kappa \\
-i\kappa & a-i\Delta \gamma +i \gamma_0
\end{array}\right)
U_u^{-1}, \nonumber \\
& = & 
 U_u (\Upsilon_{PT} + i \gamma_0 \textbf{I})]U_u^{-1}, 
\label{eq_finalgauge}
\end{eqnarray}
where $\textbf{I}$ is the identity matrix and the condition for the EP remains as earlier, $\Delta\gamma=\pm\kappa$. Apart from the above, we can also choose to gauge the system towards fully real Willis coupling such that $\text{Im}(\chi)=0$. In this case, one can see in Fig. \ref{figGauge} marked with vertical gray dashed lines in the blue areas that $\text{Im}(K_{\text{eff}}^{-1})=\text{Im}(\rho_{eff})=-i\gamma'_0$ and that $\text{Re}(K_{\text{eff}}^{-1}-\rho_{\text{eff}})=\pm2\text{Re}(\chi)$, which correspond to the EP condition for the physical effective system when transformed from $\Upsilon_R$
 \begin{eqnarray}
\textbf{M}_{\textbf{eff}}   & = & 
 U_u \left(\begin{array}{cc}
a+\Delta \gamma+i \gamma'_0 & \kappa \\
-\kappa & a-\Delta\gamma +i \gamma'_0
\end{array}\right)
U_u^{-1}, \nonumber \\
& = & 
 U_u (\Upsilon_R + i \gamma'_0 \textbf{I})]U_u^{-1}.  
\label{eq_finalgauge}
\end{eqnarray}
Thanks to the fact that the signature of $\mathcal{PT}$ symmetry breaking continues to exist even when an average loss bias is applied to ideal $\mathcal{PT}$ Hamiltonians, the EP embedded in the passive acoustic system will not cease to exist. Hence, our passive system that is described by effective constitutive relations containing the Willis coupling can be mapped onto two different forms of the ideal $\mathcal{PT}$ Hamiltonians using an unitary gauge transformation,  which conclusively unifies the resulting asymmetry in the acoustic reflections. \\
\begin{figure}[ht!]
\centering
\includegraphics[width=8.6cm]{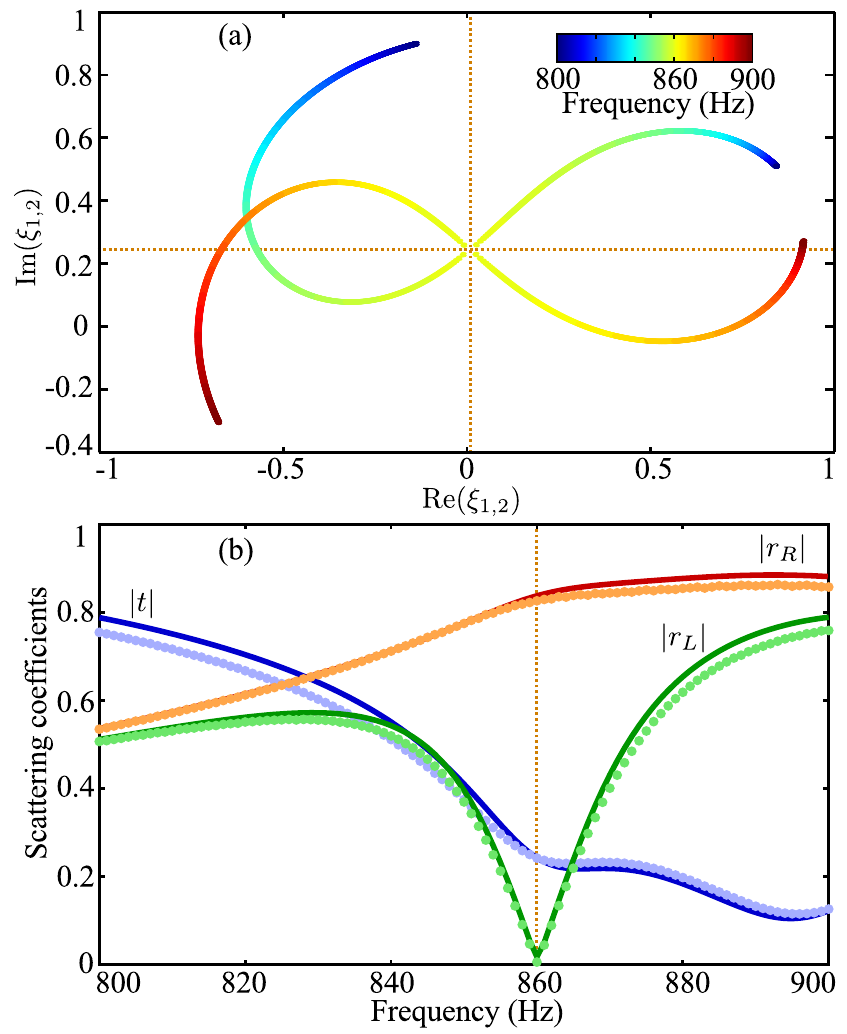}
\caption{(color online) (a) Eigenvalues spectrum of the scattering matrix. The EP that originates from the coalescence of the eigenvalues of the scattering matrix, is marked with the crossing of orange lines. (b) Theoretical (continuous curves) and experimental (dots) scattering coefficients of the sample. The EP where $r_L\rightarrow0$ is marked with an orange vertical line. \label{figEigScat}}
\end{figure}
In what follows, we show how the EP embedded in the $\textbf{M}_{\textbf{eff}}$ matrix translates into the scattering properties of the sample. The reciprocal scattering matrix that relates the amplitude of the outcoming waves to the amplitude of the incoming waves is here defined as
\begin{eqnarray}
\textbf{S} =  \left(\begin{array}{cc} t & r_R \\ r_L & t \end{array}\right). 
\label{eqscat}
\end{eqnarray} 
Its eigenvalues are $\xi_{1,2}=t\pm(r_Lr_R)^{1/2}$ and its eigenvectors can be written either as $(\sqrt{r_R},\pm\sqrt{r_L})$ or $(\pm \sqrt{r_R},\sqrt{r_L})$. 
The coalescence of the eigenstates of this scattering matrix is characterized by either $r^L=0$ or $r^R=0$ and $r^L\neq r^R$. In Fig. 1(c) we demonstrated that the eigenvalues $\zeta_{1,2}$ of the constitutive matrix coalesce at 860~Hz and agrees perfectly well with the spectral location of the EP of the associated scattering matrix as it can be seen in Fig. \ref{figEigScat}(a). The coalescence of the scattering eigenvalues at that point yields reflectionless propagation for left incidence $r_L\rightarrow0\neq r_R$ as seen in Fig. \ref{figEigScat}(b), where experimental data agree very well to theoretical predictions in asymmetrically side-loaded waveguides. We further emphasize that the size of the cell is substantially smaller than the incident wavelength $\lambda$, i.e., $L<\lambda/13$ and that the unidirectional reflectionless behavior is associated with high levels of absorption within the sample and a consequential low transmittance. For right incidence on the other hand, the wave is mostly reflected with lower level of absorption, denoting a high level of asymmetry in both the absorptions and reflections. Interestingly, the $\mathcal{PT}$ conditions for the one-dimensional scattering matrix using the generalized parity operator without considering the loss bias assume now the following form
\begin{eqnarray}
&& r_Lr_R^*e^{-4ik\delta}   = r_L^*r_R e^{4ik\delta}=1-|t|^2, \\
& & r_Lt^* +  r_L^*t e^{4ik\delta} = r_Rt^*+r_R^*t e^{-4ik\delta} =0. 
\end{eqnarray}
\\
In summary, we have demonstrated that an EP can appear in the constitutive relations of a passive acoustic Willis media, which can be mapped onto ideal $\mathcal{PT}$ Hamiltonians through a gauge transformation and an average loss bias. This EP translates into an unidirectional reflectionless propagation, which is, of primary importance for sound absorption because, if combined with coherent perfect absorption, it results in an unidirectional perfect absorber \cite{ramenazi2016}. Further, our present findings can lead to a deeper insight into the unidirectional invisibility phenomena \cite{zinlin2011,fleury2015}, altogether showing how the Willis coupling broadens the possibilities of embracing both worlds of acoustic metamaterials and $\mathcal{PT}$ symmetry physics at once to achieve unprecedented control of sound and vibrations. \\

J. C. acknowledges the support from the European Research Council (ERC) through the Starting Grant No. 714577 PHONOMETA and from the MINECO through a Ram\'on y Cajal grant (Grant No. RYC-2015-17156).
J.-P. Groby and V. Romero Garc\'ia gratefully acknowledge the support from the ANR Project METAUDIBLE No. ANR-13-BS09-0003 co-founded by ANR and FRAE and the support from  the  RFI  Le  Mans  Acoustique (R\'egion  Pays  de  la  Loire)  PavNat  project. This article is based upon work from COST Action DENORMS CA15125, supported by COST (European Cooperation in Science and Technology).


\begin{thebibliography}{48}%
\makeatletter
\providecommand \@ifxundefined [1]{%
 \@ifx{#1\undefined}
}%
\providecommand \@ifnum [1]{%
 \ifnum #1\expandafter \@firstoftwo
 \else \expandafter \@secondoftwo
 \fi
}%
\providecommand \@ifx [1]{%
 \ifx #1\expandafter \@firstoftwo
 \else \expandafter \@secondoftwo
 \fi
}%
\providecommand \natexlab [1]{#1}%
\providecommand \enquote  [1]{``#1''}%
\providecommand \bibnamefont  [1]{#1}%
\providecommand \bibfnamefont [1]{#1}%
\providecommand \citenamefont [1]{#1}%
\providecommand \href@noop [0]{\@secondoftwo}%
\providecommand \href [0]{\begingroup \@sanitize@url \@href}%
\providecommand \@href[1]{\@@startlink{#1}\@@href}%
\providecommand \@@href[1]{\endgroup#1\@@endlink}%
\providecommand \@sanitize@url [0]{\catcode `\\12\catcode `\$12\catcode
  `\&12\catcode `\#12\catcode `\^12\catcode `\_12\catcode `\%12\relax}%
\providecommand \@@startlink[1]{}%
\providecommand \@@endlink[0]{}%
\providecommand \url  [0]{\begingroup\@sanitize@url \@url }%
\providecommand \@url [1]{\endgroup\@href {#1}{\urlprefix }}%
\providecommand \urlprefix  [0]{URL }%
\providecommand \Eprint [0]{\href }%
\providecommand \doibase [0]{http://dx.doi.org/}%
\providecommand \selectlanguage [0]{\@gobble}%
\providecommand \bibinfo  [0]{\@secondoftwo}%
\providecommand \bibfield  [0]{\@secondoftwo}%
\providecommand \translation [1]{[#1]}%
\providecommand \BibitemOpen [0]{}%
\providecommand \bibitemStop [0]{}%
\providecommand \bibitemNoStop [0]{.\EOS\space}%
\providecommand \EOS [0]{\spacefactor3000\relax}%
\providecommand \BibitemShut  [1]{\csname bibitem#1\endcsname}%
\let\auto@bib@innerbib\@empty
\bibitem [{\citenamefont {Wegener}(2013)}]{WegenerScience}%
  \BibitemOpen
  \bibfield  {author} {\bibinfo {author} {\bibfnamefont {M.}~\bibnamefont
  {Wegener}},\ }\href@noop {} {\bibfield  {journal} {\bibinfo  {journal}
  {Science}\ }\textbf {\bibinfo {volume} {342}},\ \bibinfo {pages} {939}
  (\bibinfo {year} {2013})}\BibitemShut {NoStop}%
\bibitem [{\citenamefont {Milton}\ \emph {et~al.}(2006)\citenamefont {Milton},
  \citenamefont {Briane},\ and\ \citenamefont {Willis}}]{milton2006}%
  \BibitemOpen
  \bibfield  {author} {\bibinfo {author} {\bibfnamefont {G.~W.}\ \bibnamefont
  {Milton}}, \bibinfo {author} {\bibfnamefont {M.}~\bibnamefont {Briane}}, \
  and\ \bibinfo {author} {\bibfnamefont {J.~R.}\ \bibnamefont {Willis}},\
  }\href@noop {} {\bibfield  {journal} {\bibinfo  {journal} {New J. Phys.}\
  }\textbf {\bibinfo {volume} {8}},\ \bibinfo {pages} {248} (\bibinfo {year}
  {2006})}\BibitemShut {NoStop}%
\bibitem [{\citenamefont {Norris}(2008)}]{norris2008}%
  \BibitemOpen
  \bibfield  {author} {\bibinfo {author} {\bibfnamefont {A.~N.}\ \bibnamefont
  {Norris}},\ }\href@noop {} {\bibfield  {journal} {\bibinfo  {journal} {Proc.
  R. Soc. A}\ }\textbf {\bibinfo {volume} {464}},\ \bibinfo {pages} {2411}
  (\bibinfo {year} {2008})}\BibitemShut {NoStop}%
\bibitem [{\citenamefont {Torrent}\ and\ \citenamefont
  {S\'anchez-Dehesa}(2008)}]{torrent2008}%
  \BibitemOpen
  \bibfield  {author} {\bibinfo {author} {\bibfnamefont {D.}~\bibnamefont
  {Torrent}}\ and\ \bibinfo {author} {\bibfnamefont {J.}~\bibnamefont
  {S\'anchez-Dehesa}},\ }\href@noop {} {\bibfield  {journal} {\bibinfo
  {journal} {New J. Phys.}\ }\textbf {\bibinfo {volume} {10}},\ \bibinfo
  {pages} {063015} (\bibinfo {year} {2008})}\BibitemShut {NoStop}%
\bibitem [{\citenamefont {Garc\'ia-Chocano}\ \emph {et~al.}(2014)\citenamefont
  {Garc\'ia-Chocano}, \citenamefont {Christensen},\ and\ \citenamefont
  {S\'anchez-Dehesa}}]{garciachocano2014}%
  \BibitemOpen
  \bibfield  {author} {\bibinfo {author} {\bibfnamefont {V.~M.}\ \bibnamefont
  {Garc\'ia-Chocano}}, \bibinfo {author} {\bibfnamefont {J.}~\bibnamefont
  {Christensen}}, \ and\ \bibinfo {author} {\bibfnamefont {J.}~\bibnamefont
  {S\'anchez-Dehesa}},\ }\href@noop {} {\bibfield  {journal} {\bibinfo
  {journal} {Phys. Rev. Lett.}\ }\textbf {\bibinfo {volume} {112}},\ \bibinfo
  {pages} {144301} (\bibinfo {year} {2014})}\BibitemShut {NoStop}%
\bibitem [{\citenamefont {Haberman}\ and\ \citenamefont
  {Norris}(2016)}]{haberman2016}%
  \BibitemOpen
  \bibfield  {author} {\bibinfo {author} {\bibfnamefont {M.~R.}\ \bibnamefont
  {Haberman}}\ and\ \bibinfo {author} {\bibfnamefont {A.~N.}\ \bibnamefont
  {Norris}},\ }\href@noop {} {\bibfield  {journal} {\bibinfo  {journal}
  {Acoust. Today}\ }\textbf {\bibinfo {volume} {12}},\ \bibinfo {pages} {31}
  (\bibinfo {year} {2016})}\BibitemShut {NoStop}%
\bibitem [{\citenamefont {Liang}\ \emph {et~al.}(2010)\citenamefont {Liang},
  \citenamefont {Guo}, \citenamefont {Tu}, \citenamefont {Zhang},\ and\
  \citenamefont {Cheng}}]{liang2010-2}%
  \BibitemOpen
  \bibfield  {author} {\bibinfo {author} {\bibfnamefont {B.}~\bibnamefont
  {Liang}}, \bibinfo {author} {\bibfnamefont {X.~S.}\ \bibnamefont {Guo}},
  \bibinfo {author} {\bibfnamefont {J.}~\bibnamefont {Tu}}, \bibinfo {author}
  {\bibfnamefont {D.}~\bibnamefont {Zhang}}, \ and\ \bibinfo {author}
  {\bibfnamefont {J.~C.}\ \bibnamefont {Cheng}},\ }\href@noop {} {\bibfield
  {journal} {\bibinfo  {journal} {Nat. Mater.}\ }\textbf {\bibinfo {volume}
  {9}},\ \bibinfo {pages} {989} (\bibinfo {year} {2010})}\BibitemShut {NoStop}%
\bibitem [{\citenamefont {Boechler}\ \emph {et~al.}(2011)\citenamefont
  {Boechler}, \citenamefont {Theocharis},\ and\ \citenamefont
  {Daraio}}]{boechler2011}%
  \BibitemOpen
  \bibfield  {author} {\bibinfo {author} {\bibfnamefont {N.}~\bibnamefont
  {Boechler}}, \bibinfo {author} {\bibfnamefont {G.}~\bibnamefont
  {Theocharis}}, \ and\ \bibinfo {author} {\bibfnamefont {C.}~\bibnamefont
  {Daraio}},\ }\href@noop {} {\bibfield  {journal} {\bibinfo  {journal} {Nat.
  Mater.}\ }\textbf {\bibinfo {volume} {10}},\ \bibinfo {pages} {665} (\bibinfo
  {year} {2011})}\BibitemShut {NoStop}%
\bibitem [{\citenamefont {Fleury}\ \emph {et~al.}(2014)\citenamefont {Fleury},
  \citenamefont {Sounas}, \citenamefont {Sieck}, \citenamefont {Haberman},\
  and\ \citenamefont {Al{\`u}}}]{fleury2014}%
  \BibitemOpen
  \bibfield  {author} {\bibinfo {author} {\bibfnamefont {R.}~\bibnamefont
  {Fleury}}, \bibinfo {author} {\bibfnamefont {D.~L.}\ \bibnamefont {Sounas}},
  \bibinfo {author} {\bibfnamefont {C.~F.}\ \bibnamefont {Sieck}}, \bibinfo
  {author} {\bibfnamefont {M.~R.}\ \bibnamefont {Haberman}}, \ and\ \bibinfo
  {author} {\bibfnamefont {A.}~\bibnamefont {Al{\`u}}},\ }\href@noop {}
  {\bibfield  {journal} {\bibinfo  {journal} {Science}\ }\textbf {\bibinfo
  {volume} {343}},\ \bibinfo {pages} {516} (\bibinfo {year}
  {2014})}\BibitemShut {NoStop}%
\bibitem [{\citenamefont {Devaux}\ \emph {et~al.}(2015)\citenamefont {Devaux},
  \citenamefont {Tournat}, \citenamefont {Richoux},\ and\ \citenamefont
  {Pagneux}}]{devaux2015}%
  \BibitemOpen
  \bibfield  {author} {\bibinfo {author} {\bibfnamefont {T.}~\bibnamefont
  {Devaux}}, \bibinfo {author} {\bibfnamefont {V.}~\bibnamefont {Tournat}},
  \bibinfo {author} {\bibfnamefont {O.}~\bibnamefont {Richoux}}, \ and\
  \bibinfo {author} {\bibfnamefont {V.}~\bibnamefont {Pagneux}},\ }\href@noop
  {} {\bibfield  {journal} {\bibinfo  {journal} {Phys. Rev. Lett.}\ }\textbf
  {\bibinfo {volume} {115}},\ \bibinfo {pages} {234301} (\bibinfo {year}
  {2015})}\BibitemShut {NoStop}%
\bibitem [{\citenamefont {Trainiti}\ and\ \citenamefont
  {Ruzzene}(2016)}]{trainiti2016}%
  \BibitemOpen
  \bibfield  {author} {\bibinfo {author} {\bibfnamefont {G.}~\bibnamefont
  {Trainiti}}\ and\ \bibinfo {author} {\bibfnamefont {M.}~\bibnamefont
  {Ruzzene}},\ }\href@noop {} {\bibfield  {journal} {\bibinfo  {journal} {New
  J. Phys.}\ }\textbf {\bibinfo {volume} {18}},\ \bibinfo {pages} {083047}
  (\bibinfo {year} {2016})}\BibitemShut {NoStop}%
\bibitem [{\citenamefont {Merkel}\ \emph {et~al.}(2018)\citenamefont {Merkel},
  \citenamefont {Willatzen},\ and\ \citenamefont
  {Christensen}}]{merkel2018prapp}%
  \BibitemOpen
  \bibfield  {author} {\bibinfo {author} {\bibfnamefont {A.}~\bibnamefont
  {Merkel}}, \bibinfo {author} {\bibfnamefont {M.}~\bibnamefont {Willatzen}}, \
  and\ \bibinfo {author} {\bibfnamefont {J.}~\bibnamefont {Christensen}},\
  }\href@noop {} {\bibfield  {journal} {\bibinfo  {journal} {Phys. Rev.
  Applied}\ }\textbf {\bibinfo {volume} {9}},\ \bibinfo {pages} {034033}
  (\bibinfo {year} {2018})}\BibitemShut {NoStop}%
\bibitem [{\citenamefont {Cheng}\ and\ \citenamefont {Kong}(1968)}]{cheng1968}%
  \BibitemOpen
  \bibfield  {author} {\bibinfo {author} {\bibfnamefont {D.~K.}\ \bibnamefont
  {Cheng}}\ and\ \bibinfo {author} {\bibfnamefont {J.-A.}\ \bibnamefont
  {Kong}},\ }\href@noop {} {\bibfield  {journal} {\bibinfo  {journal} {Proc.
  IEEE}\ }\textbf {\bibinfo {volume} {56}},\ \bibinfo {pages} {248} (\bibinfo
  {year} {1968})}\BibitemShut {NoStop}%
\bibitem [{\citenamefont {Kong}(1972)}]{kong1972}%
  \BibitemOpen
  \bibfield  {author} {\bibinfo {author} {\bibfnamefont {J.-A.}\ \bibnamefont
  {Kong}},\ }\href@noop {} {\bibfield  {journal} {\bibinfo  {journal} {Proc.
  IEEE}\ }\textbf {\bibinfo {volume} {60}},\ \bibinfo {pages} {1036} (\bibinfo
  {year} {1972})}\BibitemShut {NoStop}%
\bibitem [{\citenamefont {Kriegler}\ \emph {et~al.}(2010)\citenamefont
  {Kriegler}, \citenamefont {Linden},\ and\ \citenamefont
  {Wegener}}]{Kriegler}%
  \BibitemOpen
  \bibfield  {author} {\bibinfo {author} {\bibfnamefont {C.~E.}\ \bibnamefont
  {Kriegler}}, \bibinfo {author} {\bibfnamefont {M.~S.}\ \bibnamefont
  {Linden}}, \ and\ \bibinfo {author} {\bibfnamefont {M.}~\bibnamefont
  {Wegener}},\ }\href@noop {} {\bibfield  {journal} {\bibinfo  {journal} {IEEE
  J. Sel. Top. Quant}\ }\textbf {\bibinfo {volume} {16}},\ \bibinfo {pages}
  {367} (\bibinfo {year} {2010})}\BibitemShut {NoStop}%
\bibitem [{\citenamefont {Marqu{\'e}s}\ \emph {et~al.}(2002)\citenamefont
  {Marqu{\'e}s}, \citenamefont {Medina},\ and\ \citenamefont
  {Rafii-El-Idrissi}}]{marques2002}%
  \BibitemOpen
  \bibfield  {author} {\bibinfo {author} {\bibfnamefont {R.}~\bibnamefont
  {Marqu{\'e}s}}, \bibinfo {author} {\bibfnamefont {F.}~\bibnamefont {Medina}},
  \ and\ \bibinfo {author} {\bibfnamefont {R.}~\bibnamefont
  {Rafii-El-Idrissi}},\ }\href@noop {} {\bibfield  {journal} {\bibinfo
  {journal} {Phys. Rev. B}\ }\textbf {\bibinfo {volume} {65}},\ \bibinfo
  {pages} {144440} (\bibinfo {year} {2002})}\BibitemShut {NoStop}%
\bibitem [{\citenamefont {Peng}\ \emph {et~al.}(2018)\citenamefont {Peng},
  \citenamefont {Wang}, \citenamefont {Yang}, \citenamefont {Chen},
  \citenamefont {Wang}, \citenamefont {Zhang},\ and\ \citenamefont
  {Chen}}]{liangpeng2018}%
  \BibitemOpen
  \bibfield  {author} {\bibinfo {author} {\bibfnamefont {L.}~\bibnamefont
  {Peng}}, \bibinfo {author} {\bibfnamefont {K.}~\bibnamefont {Wang}}, \bibinfo
  {author} {\bibfnamefont {Y.}~\bibnamefont {Yang}}, \bibinfo {author}
  {\bibfnamefont {Y.}~\bibnamefont {Chen}}, \bibinfo {author} {\bibfnamefont
  {G.}~\bibnamefont {Wang}}, \bibinfo {author} {\bibfnamefont {B.}~\bibnamefont
  {Zhang}}, \ and\ \bibinfo {author} {\bibfnamefont {H.}~\bibnamefont {Chen}},\
  }\href@noop {} {\bibfield  {journal} {\bibinfo  {journal} {Adv. Sci.}\ ,\
  \bibinfo {pages} {1700922}} (\bibinfo {year} {2018})}\BibitemShut {NoStop}%
\bibitem [{\citenamefont {Willis}(1981)}]{WillisSeminal}%
  \BibitemOpen
  \bibfield  {author} {\bibinfo {author} {\bibfnamefont {J.~R.}\ \bibnamefont
  {Willis}},\ }\href@noop {} {\bibfield  {journal} {\bibinfo  {journal} {Wave
  Motion}\ }\textbf {\bibinfo {volume} {3}},\ \bibinfo {pages} {1} (\bibinfo
  {year} {1981})}\BibitemShut {NoStop}%
\bibitem [{\citenamefont {Milton}\ and\ \citenamefont
  {Willis}(2007)}]{milton2007}%
  \BibitemOpen
  \bibfield  {author} {\bibinfo {author} {\bibfnamefont {G.~W.}\ \bibnamefont
  {Milton}}\ and\ \bibinfo {author} {\bibfnamefont {J.~R.}\ \bibnamefont
  {Willis}},\ }\href@noop {} {\bibfield  {journal} {\bibinfo  {journal} {Proc.
  R. Soc. A}\ }\textbf {\bibinfo {volume} {463}},\ \bibinfo {pages} {855}
  (\bibinfo {year} {2007})}\BibitemShut {NoStop}%
\bibitem [{\citenamefont {Torrent}\ \emph {et~al.}(2015)\citenamefont
  {Torrent}, \citenamefont {Pennec},\ and\ \citenamefont
  {Djafari-Rouhani}}]{PhysRevB.92.174110}%
  \BibitemOpen
  \bibfield  {author} {\bibinfo {author} {\bibfnamefont {D.}~\bibnamefont
  {Torrent}}, \bibinfo {author} {\bibfnamefont {Y.}~\bibnamefont {Pennec}}, \
  and\ \bibinfo {author} {\bibfnamefont {B.}~\bibnamefont {Djafari-Rouhani}},\
  }\href {\doibase 10.1103/PhysRevB.92.174110} {\bibfield  {journal} {\bibinfo
  {journal} {Phys. Rev. B}\ }\textbf {\bibinfo {volume} {92}},\ \bibinfo
  {pages} {174110} (\bibinfo {year} {2015})}\BibitemShut {NoStop}%
\bibitem [{\citenamefont {Koo}\ \emph {et~al.}(2016)\citenamefont {Koo},
  \citenamefont {Cho}, \citenamefont {Jeong},\ and\ \citenamefont
  {Park}}]{koo2016}%
  \BibitemOpen
  \bibfield  {author} {\bibinfo {author} {\bibfnamefont {S.}~\bibnamefont
  {Koo}}, \bibinfo {author} {\bibfnamefont {C.}~\bibnamefont {Cho}}, \bibinfo
  {author} {\bibfnamefont {J.-H.}\ \bibnamefont {Jeong}}, \ and\ \bibinfo
  {author} {\bibfnamefont {N.}~\bibnamefont {Park}},\ }\href@noop {} {\bibfield
   {journal} {\bibinfo  {journal} {Nat. Commun.}\ }\textbf {\bibinfo {volume}
  {7}},\ \bibinfo {pages} {13012} (\bibinfo {year} {2016})}\BibitemShut
  {NoStop}%
\bibitem [{\citenamefont {Muhlestein}\ \emph {et~al.}(2016)\citenamefont
  {Muhlestein}, \citenamefont {Sieck}, \citenamefont {Al{\`u}},\ and\
  \citenamefont {Haberman}}]{muhlestein2016}%
  \BibitemOpen
  \bibfield  {author} {\bibinfo {author} {\bibfnamefont {M.~B.}\ \bibnamefont
  {Muhlestein}}, \bibinfo {author} {\bibfnamefont {C.~F.}\ \bibnamefont
  {Sieck}}, \bibinfo {author} {\bibfnamefont {A.}~\bibnamefont {Al{\`u}}}, \
  and\ \bibinfo {author} {\bibfnamefont {M.~R.}\ \bibnamefont {Haberman}},\
  }\href@noop {} {\bibfield  {journal} {\bibinfo  {journal} {Proc. R. Soc. A}\
  }\textbf {\bibinfo {volume} {472}},\ \bibinfo {pages} {20160604} (\bibinfo
  {year} {2016})}\BibitemShut {NoStop}%
\bibitem [{\citenamefont {Muhlestein}\ \emph {et~al.}(2017)\citenamefont
  {Muhlestein}, \citenamefont {Sieck}, \citenamefont {Wilson},\ and\
  \citenamefont {Haberman}}]{muhlestein2017}%
  \BibitemOpen
  \bibfield  {author} {\bibinfo {author} {\bibfnamefont {M.~B.}\ \bibnamefont
  {Muhlestein}}, \bibinfo {author} {\bibfnamefont {C.~F.}\ \bibnamefont
  {Sieck}}, \bibinfo {author} {\bibfnamefont {P.~S.}\ \bibnamefont {Wilson}}, \
  and\ \bibinfo {author} {\bibfnamefont {M.~R.}\ \bibnamefont {Haberman}},\
  }\href@noop {} {\bibfield  {journal} {\bibinfo  {journal} {Nat. Commun.}\
  }\textbf {\bibinfo {volume} {8}},\ \bibinfo {pages} {15625} (\bibinfo {year}
  {2017})}\BibitemShut {NoStop}%
\bibitem [{\citenamefont {Sieck}\ \emph {et~al.}(2017)\citenamefont {Sieck},
  \citenamefont {Al{\`u}},\ and\ \citenamefont {Haberman}}]{sieck2017}%
  \BibitemOpen
  \bibfield  {author} {\bibinfo {author} {\bibfnamefont {C.~F.}\ \bibnamefont
  {Sieck}}, \bibinfo {author} {\bibfnamefont {A.}~\bibnamefont {Al{\`u}}}, \
  and\ \bibinfo {author} {\bibfnamefont {M.~R.}\ \bibnamefont {Haberman}},\
  }\href@noop {} {\bibfield  {journal} {\bibinfo  {journal} {Phys. Rev. B}\
  }\textbf {\bibinfo {volume} {96}},\ \bibinfo {pages} {104303} (\bibinfo
  {year} {2017})}\BibitemShut {NoStop}%
\bibitem [{\citenamefont {Quan}\ \emph {et~al.}(2018)\citenamefont {Quan},
  \citenamefont {Ra'di}, \citenamefont {Sounas},\ and\ \citenamefont
  {Al{\`u}}}]{liquan2018}%
  \BibitemOpen
  \bibfield  {author} {\bibinfo {author} {\bibfnamefont {L.}~\bibnamefont
  {Quan}}, \bibinfo {author} {\bibfnamefont {Y.}~\bibnamefont {Ra'di}},
  \bibinfo {author} {\bibfnamefont {D.}~\bibnamefont {Sounas}}, \ and\ \bibinfo
  {author} {\bibfnamefont {A.}~\bibnamefont {Al{\`u}}},\ }\href@noop {}
  {\bibfield  {journal} {\bibinfo  {journal} {Phys. Rev. Lett.}\ }\textbf
  {\bibinfo {volume} {120}},\ \bibinfo {pages} {254301} (\bibinfo {year}
  {2018})}\BibitemShut {NoStop}%
\bibitem [{\citenamefont {El-Ganainy}\ \emph {et~al.}(2018)\citenamefont
  {El-Ganainy}, \citenamefont {Makris}, \citenamefont {Khajavikhan},
  \citenamefont {Musslimani}, \citenamefont {Rotter},\ and\ \citenamefont
  {Christodoulides}}]{elganainy2018}%
  \BibitemOpen
  \bibfield  {author} {\bibinfo {author} {\bibfnamefont {R.}~\bibnamefont
  {El-Ganainy}}, \bibinfo {author} {\bibfnamefont {K.~G.}\ \bibnamefont
  {Makris}}, \bibinfo {author} {\bibfnamefont {M.}~\bibnamefont {Khajavikhan}},
  \bibinfo {author} {\bibfnamefont {Z.~H.}\ \bibnamefont {Musslimani}},
  \bibinfo {author} {\bibfnamefont {S.}~\bibnamefont {Rotter}}, \ and\ \bibinfo
  {author} {\bibfnamefont {D.~N.}\ \bibnamefont {Christodoulides}},\
  }\href@noop {} {\bibfield  {journal} {\bibinfo  {journal} {Nat. Phys.}\
  }\textbf {\bibinfo {volume} {14}},\ \bibinfo {pages} {11} (\bibinfo {year}
  {2018})}\BibitemShut {NoStop}%
\bibitem [{\citenamefont {Lin}\ \emph {et~al.}(2011)\citenamefont {Lin},
  \citenamefont {Ramenazi}, \citenamefont {Eichelkraut}, \citenamefont
  {Kottos}, \citenamefont {Cao},\ and\ \citenamefont
  {Christodoulides}}]{zinlin2011}%
  \BibitemOpen
  \bibfield  {author} {\bibinfo {author} {\bibfnamefont {Z.}~\bibnamefont
  {Lin}}, \bibinfo {author} {\bibfnamefont {H.}~\bibnamefont {Ramenazi}},
  \bibinfo {author} {\bibfnamefont {T.}~\bibnamefont {Eichelkraut}}, \bibinfo
  {author} {\bibfnamefont {T.}~\bibnamefont {Kottos}}, \bibinfo {author}
  {\bibfnamefont {H.}~\bibnamefont {Cao}}, \ and\ \bibinfo {author}
  {\bibfnamefont {D.~N.}\ \bibnamefont {Christodoulides}},\ }\href@noop {}
  {\bibfield  {journal} {\bibinfo  {journal} {Phys. Rev. Lett.}\ }\textbf
  {\bibinfo {volume} {106}},\ \bibinfo {pages} {213901} (\bibinfo {year}
  {2011})}\BibitemShut {NoStop}%
\bibitem [{\citenamefont {Ge}\ \emph {et~al.}(2012)\citenamefont {Ge},
  \citenamefont {Chong},\ and\ \citenamefont {Stone}}]{lige2012PRA}%
  \BibitemOpen
  \bibfield  {author} {\bibinfo {author} {\bibfnamefont {L.}~\bibnamefont
  {Ge}}, \bibinfo {author} {\bibfnamefont {Y.~D.}\ \bibnamefont {Chong}}, \
  and\ \bibinfo {author} {\bibfnamefont {A.~D.}\ \bibnamefont {Stone}},\
  }\href@noop {} {\bibfield  {journal} {\bibinfo  {journal} {Phys. Rev. A}\
  }\textbf {\bibinfo {volume} {85}},\ \bibinfo {pages} {023802} (\bibinfo
  {year} {2012})}\BibitemShut {NoStop}%
\bibitem [{\citenamefont {Fleury}\ \emph {et~al.}(2015)\citenamefont {Fleury},
  \citenamefont {Sounas},\ and\ \citenamefont {Al{\`u}}}]{fleury2015}%
  \BibitemOpen
  \bibfield  {author} {\bibinfo {author} {\bibfnamefont {R.}~\bibnamefont
  {Fleury}}, \bibinfo {author} {\bibfnamefont {D.}~\bibnamefont {Sounas}}, \
  and\ \bibinfo {author} {\bibfnamefont {A.}~\bibnamefont {Al{\`u}}},\
  }\href@noop {} {\bibfield  {journal} {\bibinfo  {journal} {Nat. Commun.}\
  }\textbf {\bibinfo {volume} {6}},\ \bibinfo {pages} {5905} (\bibinfo {year}
  {2015})}\BibitemShut {NoStop}%
\bibitem [{\citenamefont {Shi}\ \emph {et~al.}(2016)\citenamefont {Shi},
  \citenamefont {ans Yun~Chen}, \citenamefont {Cheng}, \citenamefont
  {Ramenazi}, \citenamefont {Wang},\ and\ \citenamefont
  {Zhang}}]{chengzhishi2016}%
  \BibitemOpen
  \bibfield  {author} {\bibinfo {author} {\bibfnamefont {C.}~\bibnamefont
  {Shi}}, \bibinfo {author} {\bibfnamefont {M.~D.}\ \bibnamefont {ans
  Yun~Chen}}, \bibinfo {author} {\bibfnamefont {L.}~\bibnamefont {Cheng}},
  \bibinfo {author} {\bibfnamefont {H.}~\bibnamefont {Ramenazi}}, \bibinfo
  {author} {\bibfnamefont {Y.}~\bibnamefont {Wang}}, \ and\ \bibinfo {author}
  {\bibfnamefont {X.}~\bibnamefont {Zhang}},\ }\href@noop {} {\bibfield
  {journal} {\bibinfo  {journal} {Nat. Commun.}\ }\textbf {\bibinfo {volume}
  {7}},\ \bibinfo {pages} {11110} (\bibinfo {year} {2016})}\BibitemShut
  {NoStop}%
\bibitem [{\citenamefont {Christensen}\ \emph {et~al.}(2016)\citenamefont
  {Christensen}, \citenamefont {Willatzen}, \citenamefont {Velasco},\ and\
  \citenamefont {Lu}}]{PhysRevLett.116.207601}%
  \BibitemOpen
  \bibfield  {author} {\bibinfo {author} {\bibfnamefont {J.}~\bibnamefont
  {Christensen}}, \bibinfo {author} {\bibfnamefont {M.}~\bibnamefont
  {Willatzen}}, \bibinfo {author} {\bibfnamefont {V.~R.}\ \bibnamefont
  {Velasco}}, \ and\ \bibinfo {author} {\bibfnamefont {M.-H.}\ \bibnamefont
  {Lu}},\ }\href@noop {} {\bibfield  {journal} {\bibinfo  {journal} {Phys. Rev.
  Lett.}\ }\textbf {\bibinfo {volume} {116}},\ \bibinfo {pages} {207601}
  (\bibinfo {year} {2016})}\BibitemShut {NoStop}%
\bibitem [{\citenamefont {Guo}\ \emph {et~al.}(2009)\citenamefont {Guo},
  \citenamefont {Salamo}, \citenamefont {Duchesne}, \citenamefont {Morandotti},
  \citenamefont {Volatier-Ravat}, \citenamefont {Aimez}, \citenamefont
  {Siviloglou},\ and\ \citenamefont {Christodoulides}}]{Aguo2009}%
  \BibitemOpen
  \bibfield  {author} {\bibinfo {author} {\bibfnamefont {A.}~\bibnamefont
  {Guo}}, \bibinfo {author} {\bibfnamefont {J.}~\bibnamefont {Salamo}},
  \bibinfo {author} {\bibfnamefont {D.}~\bibnamefont {Duchesne}}, \bibinfo
  {author} {\bibfnamefont {R.}~\bibnamefont {Morandotti}}, \bibinfo {author}
  {\bibfnamefont {M.}~\bibnamefont {Volatier-Ravat}}, \bibinfo {author}
  {\bibfnamefont {V.}~\bibnamefont {Aimez}}, \bibinfo {author} {\bibfnamefont
  {G.~A.}\ \bibnamefont {Siviloglou}}, \ and\ \bibinfo {author} {\bibfnamefont
  {D.~N.}\ \bibnamefont {Christodoulides}},\ }\href@noop {} {\bibfield
  {journal} {\bibinfo  {journal} {Phys. Rev. Lett.}\ }\textbf {\bibinfo
  {volume} {103}},\ \bibinfo {pages} {093902} (\bibinfo {year}
  {2009})}\BibitemShut {NoStop}%
\bibitem [{\citenamefont {R{\"u}ter}\ \emph {et~al.}(2010)\citenamefont
  {R{\"u}ter}, \citenamefont {Makris}, \citenamefont {El-Ganainy},
  \citenamefont {Christodoulides}, \citenamefont {Segev},\ and\ \citenamefont
  {Kip}}]{ruter2010}%
  \BibitemOpen
  \bibfield  {author} {\bibinfo {author} {\bibfnamefont {C.~E.}\ \bibnamefont
  {R{\"u}ter}}, \bibinfo {author} {\bibfnamefont {K.~G.}\ \bibnamefont
  {Makris}}, \bibinfo {author} {\bibfnamefont {R.}~\bibnamefont {El-Ganainy}},
  \bibinfo {author} {\bibfnamefont {D.~N.}\ \bibnamefont {Christodoulides}},
  \bibinfo {author} {\bibfnamefont {M.}~\bibnamefont {Segev}}, \ and\ \bibinfo
  {author} {\bibfnamefont {D.}~\bibnamefont {Kip}},\ }\href@noop {} {\bibfield
  {journal} {\bibinfo  {journal} {Nat. Phys.}\ }\textbf {\bibinfo {volume}
  {6}},\ \bibinfo {pages} {192} (\bibinfo {year} {2010})}\BibitemShut {NoStop}%
\bibitem [{\citenamefont {Kang}\ \emph {et~al.}(2013)\citenamefont {Kang},
  \citenamefont {Liu},\ and\ \citenamefont {Li}}]{minkang2013}%
  \BibitemOpen
  \bibfield  {author} {\bibinfo {author} {\bibfnamefont {M.}~\bibnamefont
  {Kang}}, \bibinfo {author} {\bibfnamefont {F.}~\bibnamefont {Liu}}, \ and\
  \bibinfo {author} {\bibfnamefont {J.}~\bibnamefont {Li}},\ }\href@noop {}
  {\bibfield  {journal} {\bibinfo  {journal} {Phys. Rev. A}\ }\textbf {\bibinfo
  {volume} {87}},\ \bibinfo {pages} {053824} (\bibinfo {year}
  {2013})}\BibitemShut {NoStop}%
\bibitem [{\citenamefont {Feng}\ \emph {et~al.}(2013)\citenamefont {Feng},
  \citenamefont {Xu}, \citenamefont {Fegadolli}, \citenamefont {Lu},
  \citenamefont {Oliveira}, \citenamefont {Almeida}, \citenamefont {Chen},\
  and\ \citenamefont {Scherer}}]{liangfeng2013}%
  \BibitemOpen
  \bibfield  {author} {\bibinfo {author} {\bibfnamefont {L.}~\bibnamefont
  {Feng}}, \bibinfo {author} {\bibfnamefont {Y.-L.}\ \bibnamefont {Xu}},
  \bibinfo {author} {\bibfnamefont {W.~S.}\ \bibnamefont {Fegadolli}}, \bibinfo
  {author} {\bibfnamefont {M.-H.}\ \bibnamefont {Lu}}, \bibinfo {author}
  {\bibfnamefont {J.~E.~B.}\ \bibnamefont {Oliveira}}, \bibinfo {author}
  {\bibfnamefont {V.~R.}\ \bibnamefont {Almeida}}, \bibinfo {author}
  {\bibfnamefont {Y.-F.}\ \bibnamefont {Chen}}, \ and\ \bibinfo {author}
  {\bibfnamefont {A.}~\bibnamefont {Scherer}},\ }\href@noop {} {\bibfield
  {journal} {\bibinfo  {journal} {Nat. Mater.}\ }\textbf {\bibinfo {volume}
  {12}},\ \bibinfo {pages} {108} (\bibinfo {year} {2013})}\BibitemShut
  {NoStop}%
\bibitem [{\citenamefont {Sun}\ \emph {et~al.}(2014)\citenamefont {Sun},
  \citenamefont {Tan}, \citenamefont {Li}, \citenamefont {Li},\ and\
  \citenamefont {Chen}}]{yongsun2014}%
  \BibitemOpen
  \bibfield  {author} {\bibinfo {author} {\bibfnamefont {Y.}~\bibnamefont
  {Sun}}, \bibinfo {author} {\bibfnamefont {W.}~\bibnamefont {Tan}}, \bibinfo
  {author} {\bibfnamefont {H.-Q.}\ \bibnamefont {Li}}, \bibinfo {author}
  {\bibfnamefont {J.}~\bibnamefont {Li}}, \ and\ \bibinfo {author}
  {\bibfnamefont {H.}~\bibnamefont {Chen}},\ }\href@noop {} {\bibfield
  {journal} {\bibinfo  {journal} {Phys. Rev. Lett.}\ }\textbf {\bibinfo
  {volume} {112}},\ \bibinfo {pages} {143903} (\bibinfo {year}
  {2014})}\BibitemShut {NoStop}%
\bibitem [{\citenamefont {Lawrence}\ \emph {et~al.}(2014)\citenamefont
  {Lawrence}, \citenamefont {Xu}, \citenamefont {Zhang}, \citenamefont {Cong},\
  and\ \citenamefont {Zhang}}]{lawrence2014}%
  \BibitemOpen
  \bibfield  {author} {\bibinfo {author} {\bibfnamefont {M.}~\bibnamefont
  {Lawrence}}, \bibinfo {author} {\bibfnamefont {N.}~\bibnamefont {Xu}},
  \bibinfo {author} {\bibfnamefont {X.}~\bibnamefont {Zhang}}, \bibinfo
  {author} {\bibfnamefont {L.}~\bibnamefont {Cong}}, \ and\ \bibinfo {author}
  {\bibfnamefont {J.~H. W. Z.~S.}\ \bibnamefont {Zhang}},\ }\href@noop {}
  {\bibfield  {journal} {\bibinfo  {journal} {Phys. Rev. Lett.}\ }\textbf
  {\bibinfo {volume} {113}},\ \bibinfo {pages} {093901} (\bibinfo {year}
  {2014})}\BibitemShut {NoStop}%
\bibitem [{\citenamefont {Wu}\ \emph {et~al.}(2014)\citenamefont {Wu},
  \citenamefont {Artoni},\ and\ \citenamefont {Rocca}}]{jinhuiwu2014}%
  \BibitemOpen
  \bibfield  {author} {\bibinfo {author} {\bibfnamefont {J.-H.}\ \bibnamefont
  {Wu}}, \bibinfo {author} {\bibfnamefont {M.}~\bibnamefont {Artoni}}, \ and\
  \bibinfo {author} {\bibfnamefont {G.~C.~L.}\ \bibnamefont {Rocca}},\
  }\href@noop {} {\bibfield  {journal} {\bibinfo  {journal} {Phys. Rev. Lett.}\
  }\textbf {\bibinfo {volume} {113}},\ \bibinfo {pages} {123004} (\bibinfo
  {year} {2014})}\BibitemShut {NoStop}%
\bibitem [{\citenamefont {Yan}\ and\ \citenamefont
  {Giebink}(2014)}]{yixinyan2014}%
  \BibitemOpen
  \bibfield  {author} {\bibinfo {author} {\bibfnamefont {Y.}~\bibnamefont
  {Yan}}\ and\ \bibinfo {author} {\bibfnamefont {N.~C.}\ \bibnamefont
  {Giebink}},\ }\href@noop {} {\bibfield  {journal} {\bibinfo  {journal} {Adv.
  Optical Mat.}\ }\textbf {\bibinfo {volume} {2}},\ \bibinfo {pages} {423}
  (\bibinfo {year} {2014})}\BibitemShut {NoStop}%
\bibitem [{\citenamefont {Liu}\ \emph {et~al.}(2018)\citenamefont {Liu},
  \citenamefont {Zhu}, \citenamefont {Liang},\ and\ \citenamefont
  {Zhu}}]{tuoliu2018}%
  \BibitemOpen
  \bibfield  {author} {\bibinfo {author} {\bibfnamefont {T.}~\bibnamefont
  {Liu}}, \bibinfo {author} {\bibfnamefont {X.}~\bibnamefont {Zhu}}, \bibinfo
  {author} {\bibfnamefont {F.~C.~S.}\ \bibnamefont {Liang}}, \ and\ \bibinfo
  {author} {\bibfnamefont {J.}~\bibnamefont {Zhu}},\ }\href@noop {} {\bibfield
  {journal} {\bibinfo  {journal} {Phys. Rev. Lett.}\ }\textbf {\bibinfo
  {volume} {120}},\ \bibinfo {pages} {124502} (\bibinfo {year}
  {2018})}\BibitemShut {NoStop}%
\bibitem [{\citenamefont {Merkel}\ \emph {et~al.}(2015)\citenamefont {Merkel},
  \citenamefont {Theocharis}, \citenamefont {Richoux}, \citenamefont
  {Romero-Garc\`ia},\ and\ \citenamefont {Pagneux}}]{merkel2015APL}%
  \BibitemOpen
  \bibfield  {author} {\bibinfo {author} {\bibfnamefont {A.}~\bibnamefont
  {Merkel}}, \bibinfo {author} {\bibfnamefont {G.}~\bibnamefont {Theocharis}},
  \bibinfo {author} {\bibfnamefont {O.}~\bibnamefont {Richoux}}, \bibinfo
  {author} {\bibfnamefont {V.}~\bibnamefont {Romero-Garc\`ia}}, \ and\ \bibinfo
  {author} {\bibfnamefont {V.}~\bibnamefont {Pagneux}},\ }\href@noop {}
  {\bibfield  {journal} {\bibinfo  {journal} {Appl. Phys. Lett.}\ }\textbf
  {\bibinfo {volume} {107}},\ \bibinfo {pages} {244102} (\bibinfo {year}
  {2015})}\BibitemShut {NoStop}%
\bibitem [{\citenamefont {Jim\'enez}\ \emph {et~al.}(2017)\citenamefont
  {Jim\'enez}, \citenamefont {Romero-Grac\'ia}, \citenamefont {Pagneux},\ and\
  \citenamefont {Groby}}]{jimenez2017}%
  \BibitemOpen
  \bibfield  {author} {\bibinfo {author} {\bibfnamefont {N.}~\bibnamefont
  {Jim\'enez}}, \bibinfo {author} {\bibfnamefont {V.}~\bibnamefont
  {Romero-Grac\'ia}}, \bibinfo {author} {\bibfnamefont {V.}~\bibnamefont
  {Pagneux}}, \ and\ \bibinfo {author} {\bibfnamefont {J.-P.}\ \bibnamefont
  {Groby}},\ }\href@noop {} {\bibfield  {journal} {\bibinfo  {journal} {Sci.
  Rep.}\ }\textbf {\bibinfo {volume} {7}},\ \bibinfo {pages} {13595} (\bibinfo
  {year} {2017})}\BibitemShut {NoStop}%
\bibitem [{\citenamefont {Bender}\ \emph {et~al.}(2002)\citenamefont {Bender},
  \citenamefont {Brody},\ and\ \citenamefont {Jones}}]{bender2002}%
  \BibitemOpen
  \bibfield  {author} {\bibinfo {author} {\bibfnamefont {C.~M.}\ \bibnamefont
  {Bender}}, \bibinfo {author} {\bibfnamefont {D.~C.}\ \bibnamefont {Brody}}, \
  and\ \bibinfo {author} {\bibfnamefont {H.~F.}\ \bibnamefont {Jones}},\
  }\href@noop {} {\bibfield  {journal} {\bibinfo  {journal} {Phys. Rev. Lett.}\
  }\textbf {\bibinfo {volume} {89}},\ \bibinfo {pages} {270401} (\bibinfo
  {year} {2002})}\BibitemShut {NoStop}%
\bibitem [{\citenamefont {Mostafazadeh}(2003)}]{mostafazadeh2003}%
  \BibitemOpen
  \bibfield  {author} {\bibinfo {author} {\bibfnamefont {A.}~\bibnamefont
  {Mostafazadeh}},\ }\href@noop {} {\bibfield  {journal} {\bibinfo  {journal}
  {J. Phys. A: Math. Gen.}\ }\textbf {\bibinfo {volume} {36}},\ \bibinfo
  {pages} {7081} (\bibinfo {year} {2003})}\BibitemShut {NoStop}%
\bibitem [{\citenamefont {Ding}\ \emph {et~al.}(2016)\citenamefont {Ding},
  \citenamefont {Ma}, \citenamefont {Xiao}, \citenamefont {Zhang},\ and\
  \citenamefont {Chan}}]{kunding2016}%
  \BibitemOpen
  \bibfield  {author} {\bibinfo {author} {\bibfnamefont {K.}~\bibnamefont
  {Ding}}, \bibinfo {author} {\bibfnamefont {G.}~\bibnamefont {Ma}}, \bibinfo
  {author} {\bibfnamefont {M.}~\bibnamefont {Xiao}}, \bibinfo {author}
  {\bibfnamefont {Z.~Q.}\ \bibnamefont {Zhang}}, \ and\ \bibinfo {author}
  {\bibfnamefont {C.~T.}\ \bibnamefont {Chan}},\ }\href@noop {} {\bibfield
  {journal} {\bibinfo  {journal} {Phys. Rev. X}\ }\textbf {\bibinfo {volume}
  {6}},\ \bibinfo {pages} {021007} (\bibinfo {year} {2016})}\BibitemShut
  {NoStop}%
\bibitem [{\citenamefont {Gear}\ \emph {et~al.}(2017)\citenamefont {Gear},
  \citenamefont {Sun}, \citenamefont {Xiao}, \citenamefont {Zhang},
  \citenamefont {Fitzgerald}, \citenamefont {Rotter}, \citenamefont {Chen}, ,\
  and\ \citenamefont {Li}}]{gear2017}%
  \BibitemOpen
  \bibfield  {author} {\bibinfo {author} {\bibfnamefont {J.}~\bibnamefont
  {Gear}}, \bibinfo {author} {\bibfnamefont {Y.}~\bibnamefont {Sun}}, \bibinfo
  {author} {\bibfnamefont {S.}~\bibnamefont {Xiao}}, \bibinfo {author}
  {\bibfnamefont {L.}~\bibnamefont {Zhang}}, \bibinfo {author} {\bibfnamefont
  {R.}~\bibnamefont {Fitzgerald}}, \bibinfo {author} {\bibfnamefont
  {S.}~\bibnamefont {Rotter}}, \bibinfo {author} {\bibfnamefont
  {H.}~\bibnamefont {Chen}}, , \ and\ \bibinfo {author} {\bibfnamefont
  {J.}~\bibnamefont {Li}},\ }\href@noop {} {\bibfield  {journal} {\bibinfo
  {journal} {New J. Phys.}\ }\textbf {\bibinfo {volume} {19}},\ \bibinfo
  {pages} {123041} (\bibinfo {year} {2017})}\BibitemShut {NoStop}%
\bibitem [{\citenamefont {Cannata}\ \emph {et~al.}(2007)\citenamefont
  {Cannata}, \citenamefont {Dedonder},\ and\ \citenamefont
  {Ventura}}]{cannata2007}%
  \BibitemOpen
  \bibfield  {author} {\bibinfo {author} {\bibfnamefont {F.}~\bibnamefont
  {Cannata}}, \bibinfo {author} {\bibfnamefont {J.-P.}\ \bibnamefont
  {Dedonder}}, \ and\ \bibinfo {author} {\bibfnamefont {A.}~\bibnamefont
  {Ventura}},\ }\href@noop {} {\bibfield  {journal} {\bibinfo  {journal}
  {Annals of Physics}\ }\textbf {\bibinfo {volume} {322}},\ \bibinfo {pages}
  {397} (\bibinfo {year} {2007})}\BibitemShut {NoStop}%
\bibitem [{\citenamefont {Ramenazi}\ \emph {et~al.}(2016)\citenamefont
  {Ramenazi}, \citenamefont {Wang}, \citenamefont {Yablonovitch},\ and\
  \citenamefont {Zhang}}]{ramenazi2016}%
  \BibitemOpen
  \bibfield  {author} {\bibinfo {author} {\bibfnamefont {H.}~\bibnamefont
  {Ramenazi}}, \bibinfo {author} {\bibfnamefont {Y.}~\bibnamefont {Wang}},
  \bibinfo {author} {\bibfnamefont {E.}~\bibnamefont {Yablonovitch}}, \ and\
  \bibinfo {author} {\bibfnamefont {X.}~\bibnamefont {Zhang}},\ }\href@noop {}
  {\bibfield  {journal} {\bibinfo  {journal} {IEEE J. Sel. Topics Quant.
  Mech.}\ }\textbf {\bibinfo {volume} {22}},\ \bibinfo {pages} {5000706}
  (\bibinfo {year} {2016})}\BibitemShut {NoStop}%
\end{thebibliography}
\end{document}